\documentclass{PoS}
\usepackage[latin1]{inputenc}
\usepackage{epsfig}
\usepackage{amsmath}
\usepackage{amssymb}
\usepackage{amsbsy}
\usepackage{amsfonts}
\usepackage{graphics}
\usepackage{latexsym}
\usepackage{mathrsfs}
\usepackage{dcolumn}
\usepackage{wasysym}
\usepackage{wrapfig}

\newcommand{\bc}{\begin{center}}
\newcommand{\ec}{\end{center}}
\def\kreis{\raise0.85pt\hbox{$\scriptstyle\bigcirc$}}
\def\vollk{\lower0.85pt\hbox{\Large $\bullet$}}

\vspace*{-2.5cm}

\title{\flushleft{\normalsize \tt  DESY 19-217}
\vspace*{0.5cm}\\
\LARGE Does confinement imply CP invariance of the strong interactions?} 

\ShortTitle{Strong CP problem}

\author{\large Y. Nakamura$^\dagger$ and \speaker{G. Schierholz}$^{\ddagger\,}$\\ \\ 
$^\dagger$ RIKEN Center for Computational Science, Kobe, Hyogo
650-0047, Japan\\ 
$^\ddagger$ Deutsches Elektronen-Synchrotron DESY, 22603 Hamburg, Germany\\
\hspace*{0.01cm}E-mail: \email{nakamura@riken.jp},\, \email{gerrit.schierholz@desy.de}}

\abstract{The strong coupling constant $1/g^2$ and the vacuum angle $\theta$ of the SU(3) Yang-Mills theory are investigated in the infrared limit under the renormalization group flow. It is shown that the theory has an infrared attractive fixed point at $1/g^2 = \theta = \,0$, which leads to linear confinement and naturally solves the strong CP problem. In particular, any initial value of $\theta \neq 0$ is found to be driven to $\theta = 0$ at macroscopic distances, where quarks and gluons freeze into hadrons by the confinement mechanism.}

\FullConference{37th International Symposium on Lattice Field Theory - Lattice2019\\
		16-22 June 2019\\
		Wuhan, China}

\begin{document}

\section{Introduction}

Undoubtedly, the two most profound unsolved problems of the strong interactions are color confinement and CP invariance. While we have strong numerical evidence for color confinement, and some understanding of the dynamical mechanism that drives it, though no proof, we lack any compelling reason why CP is conserved in the strong interactions. Recall that the ground state of the theory in Euclidean space-time is the $\theta$ vacuum, arising from the CP violating term in the action
\begin{equation}
S_\theta = i\, \theta\, Q\,, \quad Q = \frac{1}{32\pi^2}\, \int d^4x\; F_{\mu\nu}^a  \tilde{F}_{\mu\nu}^a\, \in\, \mathbb{Z}\,,
\label{charge}
\end{equation}
the so-called $\theta$ term, where $Q$ is the topological charge.
A nonvanishing value of $\theta$ would result, for example, in an electric dipole moment $d_n$ of the neutron. Current experimental limits on $|d_n|$, combined with lattice calculations of $d_n/\theta$, lead to the upper bound $\displaystyle |\theta| \lesssim 7.4 \times 10^{-11}$~\cite{Guo:2015tla}. This anomalously small number is referred to as the strong CP problem. 

A popular proposal for the resolution of the strong CP problem is the Peccei-Quinn model~\cite{Peccei:1977hh}. In this model the CP violating angle $\theta$ is shifted to zero at the expense of introducing a hitherto unknown particle, the axion, thus making the theory independent of $\theta$. Other authors~\cite{Knizhnik:1984kn} have argued that the theory has an infrared attractive fixed point at $\theta = 0$, with reference to the integral quantum Hall effect~\cite{Pruisken:1984ni}, which would drive any initial value of $\theta \neq 0$ to $\theta = 0 \pmod{2\pi}$ at macroscopic length scales. Yet other models suggest that the theory does not confine for nonvanishing $\theta$. So, for example, the dual superconductor model of confinement~\cite{Kronfeld:1987ri}. In this model the monopoles acquire a color-electric charge~\cite{Witten:1979ey}, $e_m = \theta/2\pi$, which would drive the theory into a Higgs or Coulomb phase.

The idea~\cite{Knizhnik:1984kn} that the vacuum angle $\theta$ is scale dependent, like any other bare parameter of the Lagrangian, and flows to zero in the infrared limit, appears to us the most natural solution of the strong CP problem. It requires renormalization group techniques to prove it. Early investigations~\cite{Reuter:1996be} of the Yang-Mills theory using truncated renormalization group transformations show that indeed $\theta = 0$ in the infrared limit, provided the theory confines, culminating in $1/g^2 = 0$. In this work we seek a lattice solution of the problem.

We may restrict ourselves to the SU(3) Yang-Mills theory. Quarks are not expected to change the qualitative picture, as long as they are massive. In the infrared limit quarks are expected to assume masses of several hundred MeV and above. The effective Lagrangian at renormalization scale $\mu$ reads 
\begin{equation}
  \mathcal{L}(\mu) = \frac{1}{4\,g^2(\mu)}\, F_{\mu\nu}^a  F_{\mu\nu}^a \,+\, i\, \theta(\mu)\, \frac{1}{32\pi^2}\, F_{\mu\nu}^a  \tilde{F}_{\mu\nu}^a \,,
  \label{lagrangian}
\end{equation}
where each value of $\theta$ defines a different vacuum.
The static properties of the theory are revealed by the running coupling constant $g^2(\mu)$ and CP violating angle $\theta(\mu)$ in the infrared limit $\mu \rightarrow 0$. All we are left with in this limit, where quarks and gluons freeze into hadrons, are tree-level diagrams. To obtain $\mathcal{L}(\mu^\prime)$ for $\mu^\prime < \mu$ we need to integrate out the gauge fields of virtualities between $\mu^\prime$ and $\mu$. This is accomplished by the gradient flow~\cite{Luscher:2010iy}, which can be interpreted~\cite{Luscher:2013vga,Makino:2018rys} as a particular realization of the renormalization group flow.

\section{The gradient flow}

The gradient flow describes the evolution of fields as a function of flow time $t$, which can be identified with the renormalization scale $\mu = 1/\sqrt{8\,t}$ for $t \gg 0$. The flow of SU(3) gauge fields $B_\mu(t,x) = B_\mu^{\,a}(t,x)\,T^a$ is defined by~\cite{Luscher:2010iy}
\begin{equation}
\partial_{\,t}\,B_\mu = D_\nu \, G_{\mu\nu} \,, \quad G_{\mu\nu} = \partial_\mu\,B_\nu -\partial_\nu\,B_\mu + [B_\mu\, B_\nu] \,, \quad D_\mu\, \cdot = \partial_\mu \cdot + \,[B_\mu,\cdot]
\label{gflow}
\end{equation}
with the condition $B_\mu(t=0,x) = A_\mu(x)$, where $A_\mu(x)$ is the gauge field of the initial configuration. On the lattice this leads to the effective action at flow time $t$~\cite{Luscher:2009eq} 
\begin{equation}
S(B_t) = S(A) + \frac{16\, g^2}{3\, a^2} \int_0^t\! d\tau \; S(B_\tau) \,,
\label{effact}
\end{equation}
where $B_t$ stands for $B_\mu(t,x)$, and $a$ is the lattice spacing. In (\ref{effact}) the gauge fields of virtualities $\geq 1/\sqrt{8\,t}$ can be considered to be integrated out. It is expected that the physics is left unchanged under the gradient flow. In the SU(3) Yang-Mills theory this is known to be the case for the topological susceptibility $\chi_t = \langle Q^2\rangle/V$, $V$ being the space-time volume, which we confirm.


\begin{figure}[!b]
  \begin{center}
    \epsfig{file=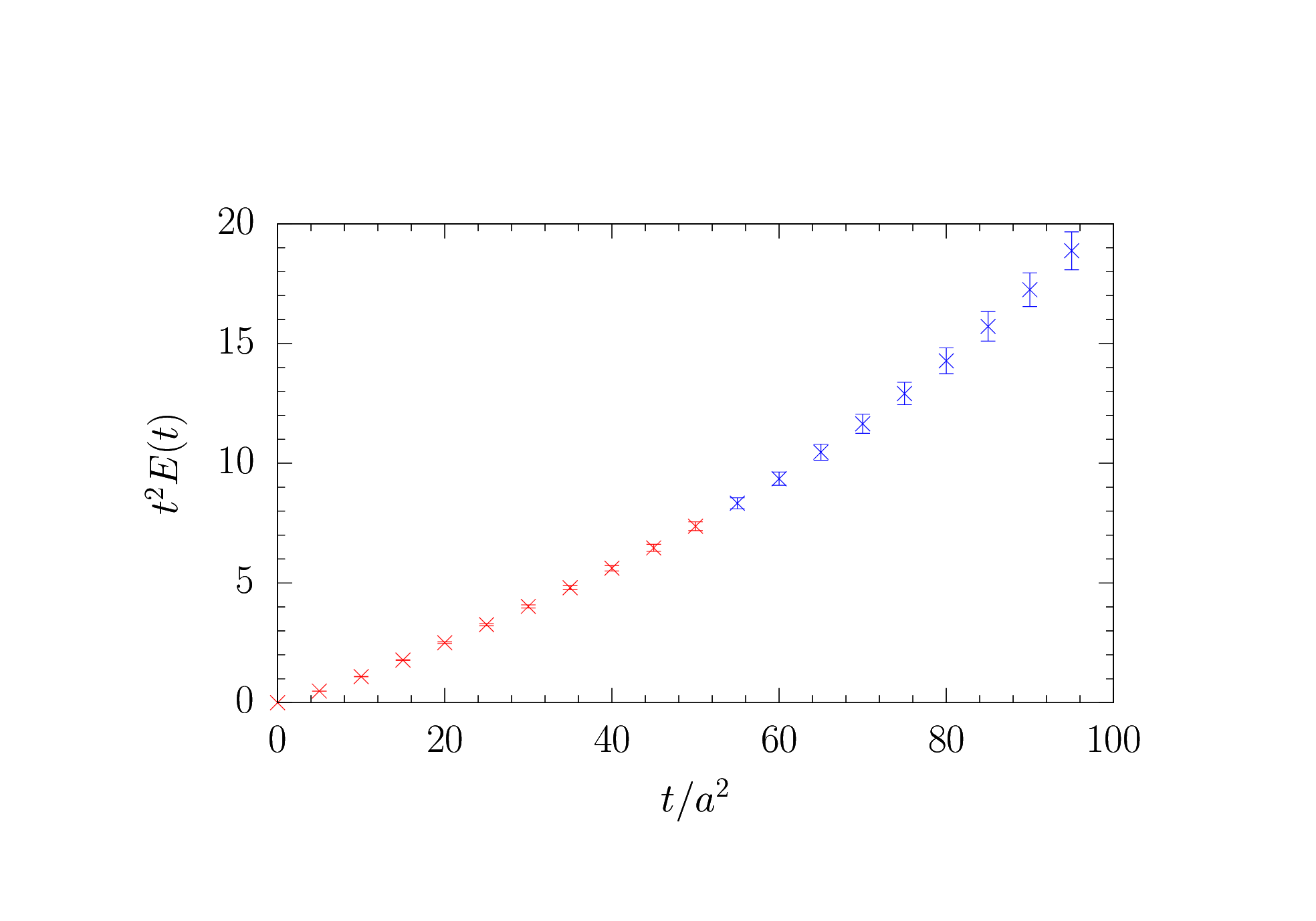,width=7.25cm,clip=}\hspace*{0.0cm}
    \epsfig{file=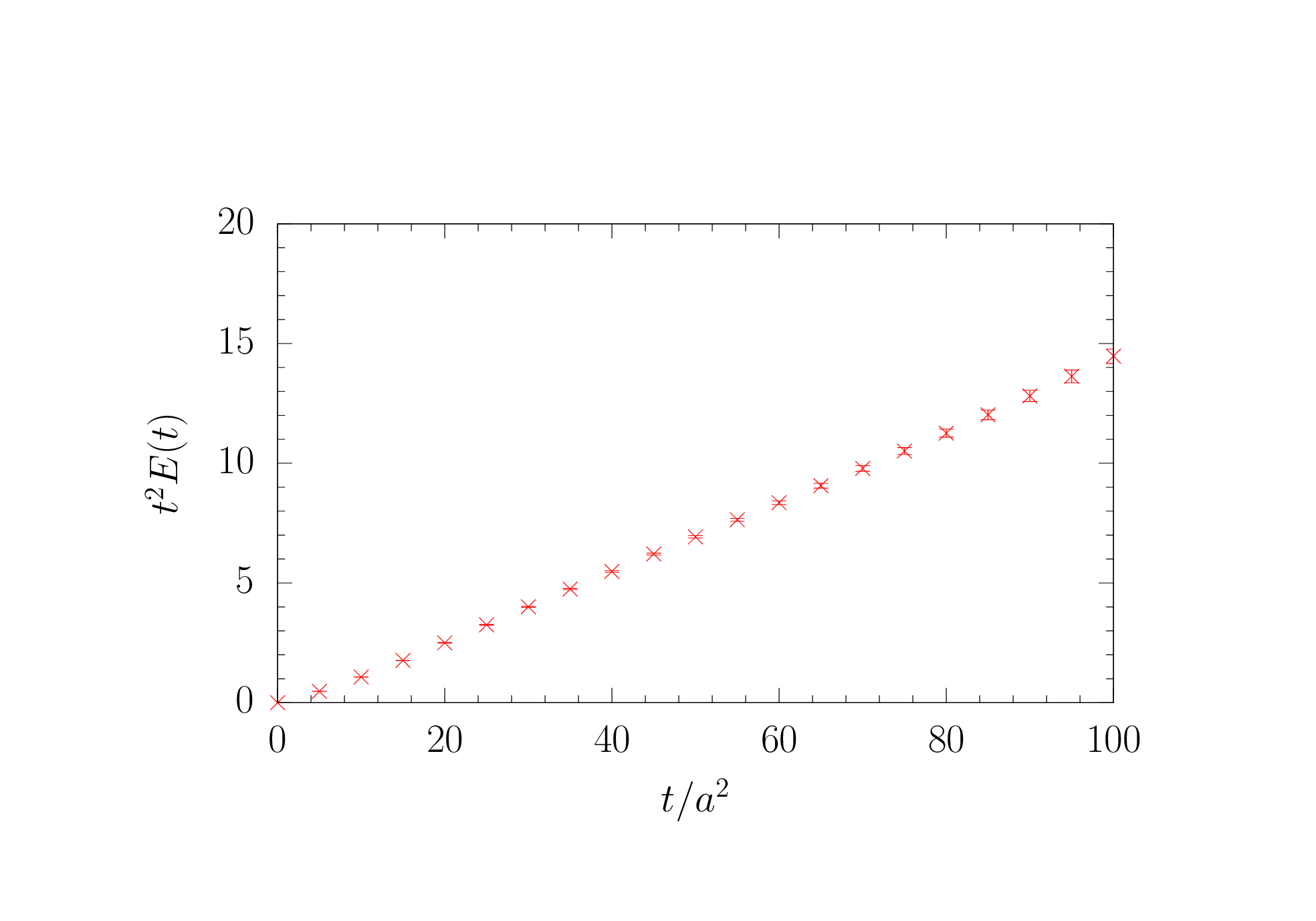,width=7.25cm,clip=}
  \end{center}
  \vspace*{-0.5cm}
\caption{The dimensionless quantity $t^2 E(t)$ as a function of $t/a^2$ on the $16^4$ lattice (left panel) and the $24^2$ lattice (right panel). At this point the topological sectors have been averaged over, corresponding to $\theta=0$.}
\label{fig1}
\end{figure}

The lattice calculation proceeds in two steps, the simulation of ensembles of gauge field configurations, to be followed by gradient flow transformations. So far we have simulated the SU(3) Yang-Mills theory on $16^4$ and $24^4$ lattices at a single value of $\beta = 6/g^2 = 6.0$, using the plaquette action
\begin{equation}
S = \beta \sum_{x,\,\mu < \nu} \Big( 1 - \frac{1}{3}\, {\rm Re}\, {\rm Tr}\; U_{\mu\nu}(x)\Big)\,.
\end{equation}
The lattice spacing at this value of $\beta$ is $a = 0.082(2) \, \mbox{fm}$, taking $\sqrt{t_0}=0.146(4)\,\mbox{fm}$ to set the scale, with $t_0$ being defined by $t_0^2\, E(t_0)=0.3$. See, for example,~\cite{Bornyakov:2015eaa}. Currently we have $4000$ uncorrelated configurations on the $16^4$ lattice and $5000$ configurations on the $24^4$ lattice at our disposal. Work on larger lattices is in progress. For our problem it is sufficient to compute the flow of the action density
\begin{equation}
  E = \frac{1}{4} G_{\mu\nu}^a  G_{\mu\nu}^a \,,
  \label{actiondensity}
\end{equation}
where $G_{\mu\nu}^a$ is the lattice version of the field tensor stated in (\ref{gflow}), and the topological charge $Q$, as we shall see. 

We consider flow times up to $t/a^2 = 100$, which corresponds to an infrared cut-off of $\mu \approx 100 \, \mbox{MeV}$. In Fig.~\ref{fig1} we show $t^2 E(t)$ as a function of $t/a^2$ on our two volumes. While both results agree for $t/a^2 \lesssim 40$, we observe finite volume effects at larger times on the $16^4$ lattice. That is not surprising, as the action density (\ref{actiondensity}) begins to get `smeared' over the whole volume in this case.

\vspace*{-0.25cm}

\section{Running coupling and linear confinement}


We consider the case $\theta=0$ first. The Yang-Mills gradient flow defines a running coupling constant at renormalization scale $\mu$ in the gradient flow scheme, 
\begin{equation}
  g^2_{GF}(\mu) = \frac{16 \pi^2}{3}\,t^2 E(t)\,.
  \label{rcGF}
\end{equation}
On the larger lattice, Fig.~\ref{fig1} (right panel), we find $t^2 E(t)$ to be a strictly linear function of $t$ for $t/a^2 \gtrsim 1$. This implies $g^2_{\rm GF}(\mu)  \propto 1/a^2\mu^2$ for $\mu \lesssim 1 \, \mbox{GeV}$, which results in the gradient flow beta function
\begin{equation}
    \frac{\partial\, \alpha_{GF}}{\partial\, \ln \, \mu} = -\,  2\; \frac{\partial\, \alpha_{GF}}{\partial\, \ln \, t} \equiv \beta_{GF}(\alpha_{\rm GF})\\
    = -\, 2 \, \alpha_{GF}(\mu)\,, \quad \alpha_{\rm GF} = \frac{g^2_{GF}}{4\,\pi}\,.
    \label{rgGF}
\end{equation}
The renormalization group equation (\ref{rgGF}) has the solution
\begin{equation}
  \frac{\Lambda_{GF}}{\mu} = \exp\left\{-\int_1^{\alpha_{GF}} \! d\alpha  \, \frac{1}{\beta_{GF}(\alpha)}\right\} = \sqrt{\alpha_{GF}}\,.
  \end{equation}
To make contact with phenomenology we need to transform the gradient flow coupling (\ref{rcGF}) to an appropriate scheme. Such a scheme is the $V$ scheme~\cite{Schroder:1998vy}. In that scheme 
\begin{equation}
    \frac{\Lambda_V}{\mu} = \exp\left\{-\int_1^{\alpha_V} \! d\alpha  \, \frac{1}{\beta_V(\alpha)}\right\}
    = \frac{\Lambda_V}{\Lambda_{GF}}\, \exp\left\{-\int_1^{\alpha_{GF}} \! d\alpha  \, \frac{1}{\beta_{GF}(\alpha)}\right\}\,.
\end{equation}
A solution of this equation is $\beta_V=-2\, \alpha_V$ with
$\displaystyle \alpha_V = (\Lambda_V/\Lambda_{GF})^2\, \alpha_{GF} = \Lambda_V^2/\mu^2 = g_V^2/4\pi$.
The ratio $\displaystyle \Lambda_V/\Lambda_{GF} = (\Lambda_V/\Lambda_{\overline{MS}})\times (\Lambda_{\overline{MS}}/\Lambda_{GF})$ is known~\cite{Luscher:2010iy,Schroder:1998vy} to be $0.731$, which finally leads to $\displaystyle \alpha_V = 0.534\, \alpha_{GF}$.

As we are left with tree-level diagrams only in the infrared regime, the long-range static potential is simply given by the exchange of a single (dressed) gluon, which reads
\begin{equation}
  V(r) = \frac{1}{(2\pi)^3} \int d^3\mathbf{q} \; e^{i\,\mathbf{q r}} \; \frac{4}{3}\, \frac{\alpha_V(q)}{\mathbf{q}^2 + i 0} \equiv \sigma \, r\,, \quad \sigma = \frac{2}{3}\, \mu^2\, \alpha_V = \frac{2}{3}\, \Lambda_V^2 \,.
\label{pot}\end{equation}
\begin{wrapfigure}[12]{r}{0.51\linewidth} \begin{center}
    \vspace*{-0.75cm}
    \epsfig{file=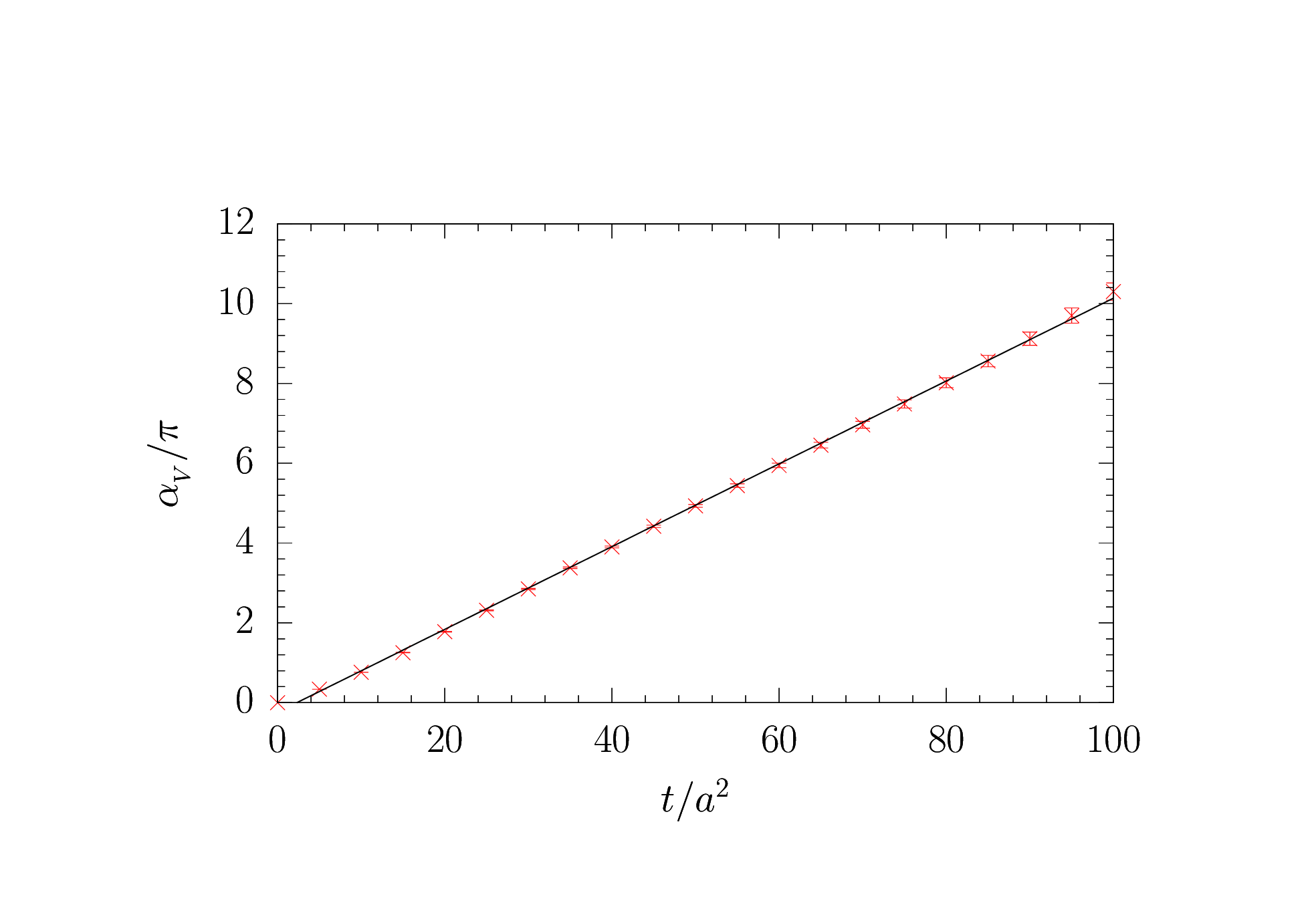,width=7.25cm,clip=}
      \end{center}
  \vspace*{-0.5cm}
  \caption{The running $\alpha_V/\pi$ as a function of $t/a^2$ on the $24^4$ lattice, together with a linear fit.}
  \label{fig2}
\end{wrapfigure}
In Fig.~\ref{fig2} we show $\alpha_V/\pi$ on the $24^4$ lattice as a function of flow time $t$, together with a linear fit of the form $\displaystyle \alpha_V/\pi = (12/\pi)\,\sigma\,t = (3/2\,\pi\,\mu^2)\,\sigma$ to the data. The result of the fit is $\sqrt{\sigma} = 396(11)\,\mbox{MeV}$, which leads to $\Lambda_V = 485(13) \,\mbox{MeV}$ and $\Lambda_{\overline{MS}} = 303(9) \,\mbox{MeV}$, respectively. Both results are in good agreement with direct lattice calculations and phenomenology. The derivation (\ref{pot}) should not be confused with attempts to read off the potential from the gluon propagator in the continuum theory.


\vspace*{-0.25cm}

\begin{figure}[h]
  \vspace*{-0.25cm}
  \begin{center}
    \epsfig{file=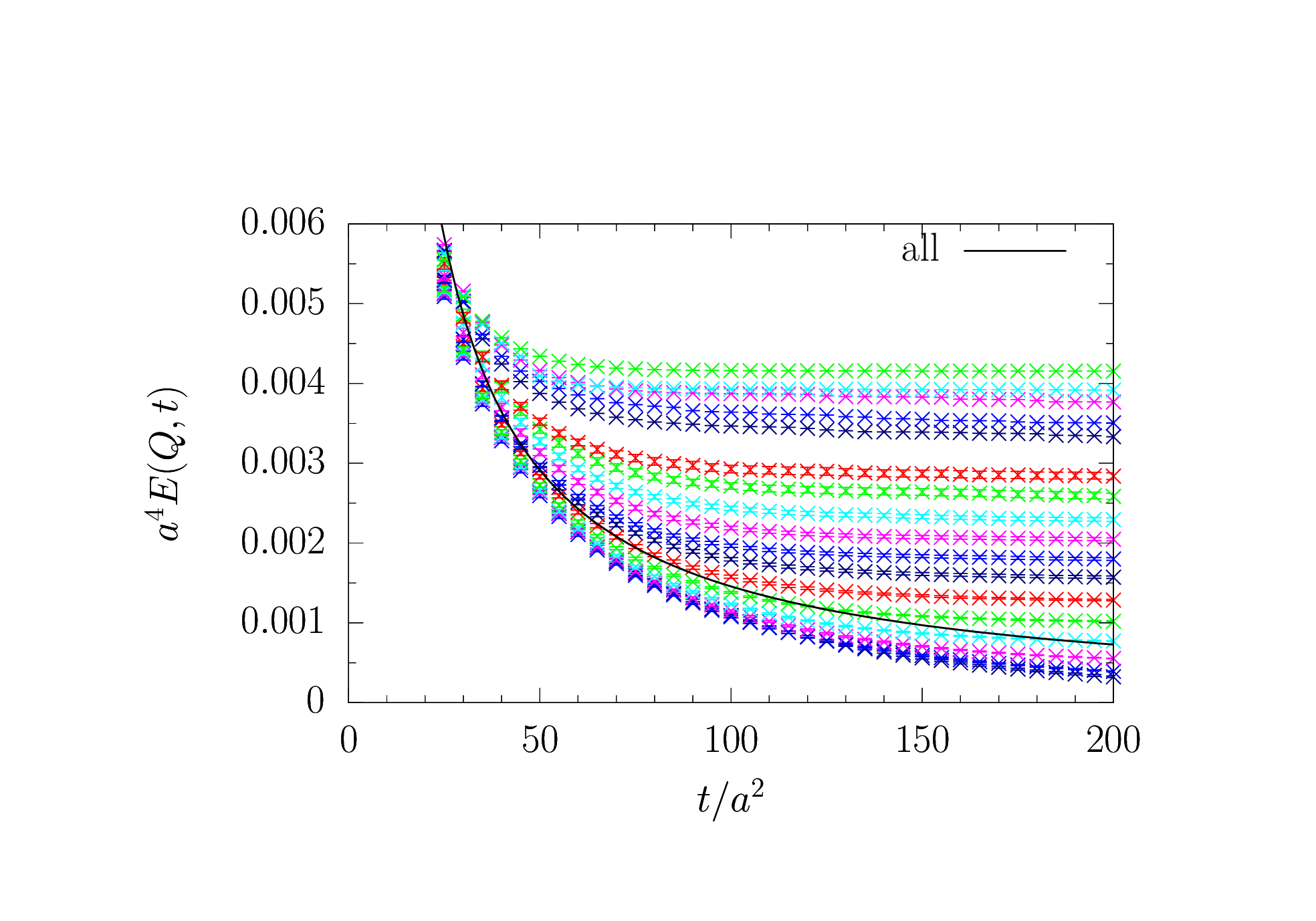,width=7.5cm,clip=}
    \epsfig{file=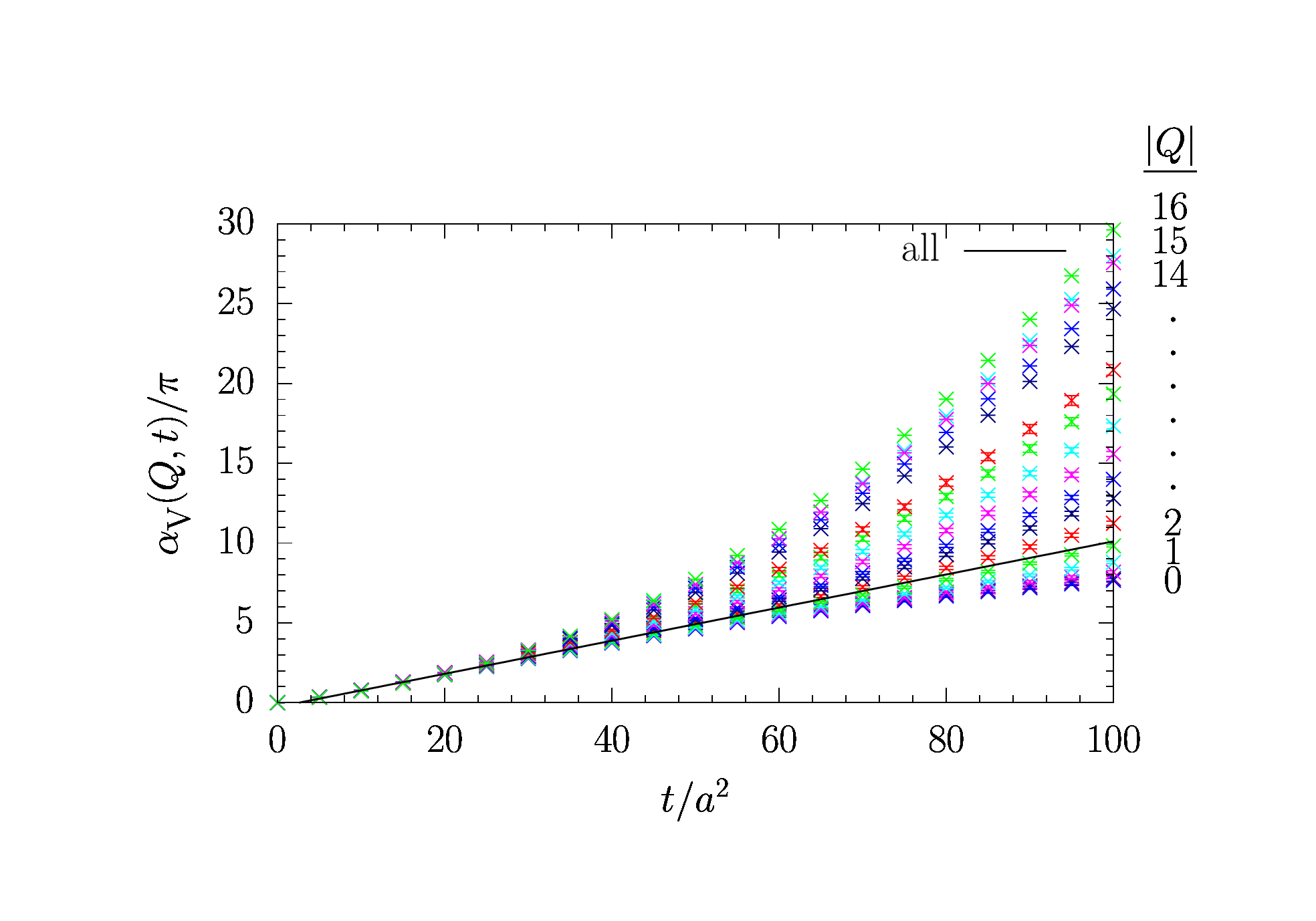,width=7.5cm,clip=}
  \end{center}
  \vspace*{-0.5cm}
  \caption{The action density $a^4 E$ (left panel) and running coupling $\alpha_V/\pi$ (right panel) broken down by the topological charge $|Q|$ on the $24^4$ lattice. $|Q|$ runs from $0$ to $16$ from bottom to top. The solid curves stand for the statistical average over all values of $Q$.}
  \label{fig3}
\end{figure}

\section{Renormalization group flow: $\mathbf{\theta}$ against $\mathbf{1/g^2}$}

It is known for a long time, albeit from `cooling', that the action density $E$ tends to multiples of the classical instanton action in the infrared regime~\cite{Ilgenfritz:1985dz}. In Fig.~\ref{fig3} we show the action density $E$ (left panel) from the gradient flow, together with the associated running coupling $\alpha_V/\pi$ (right panel),
\begin{wrapfigure}[12]{r}{0.51\linewidth} \begin{center}
  \vspace*{-0.6cm}
  \epsfig{file=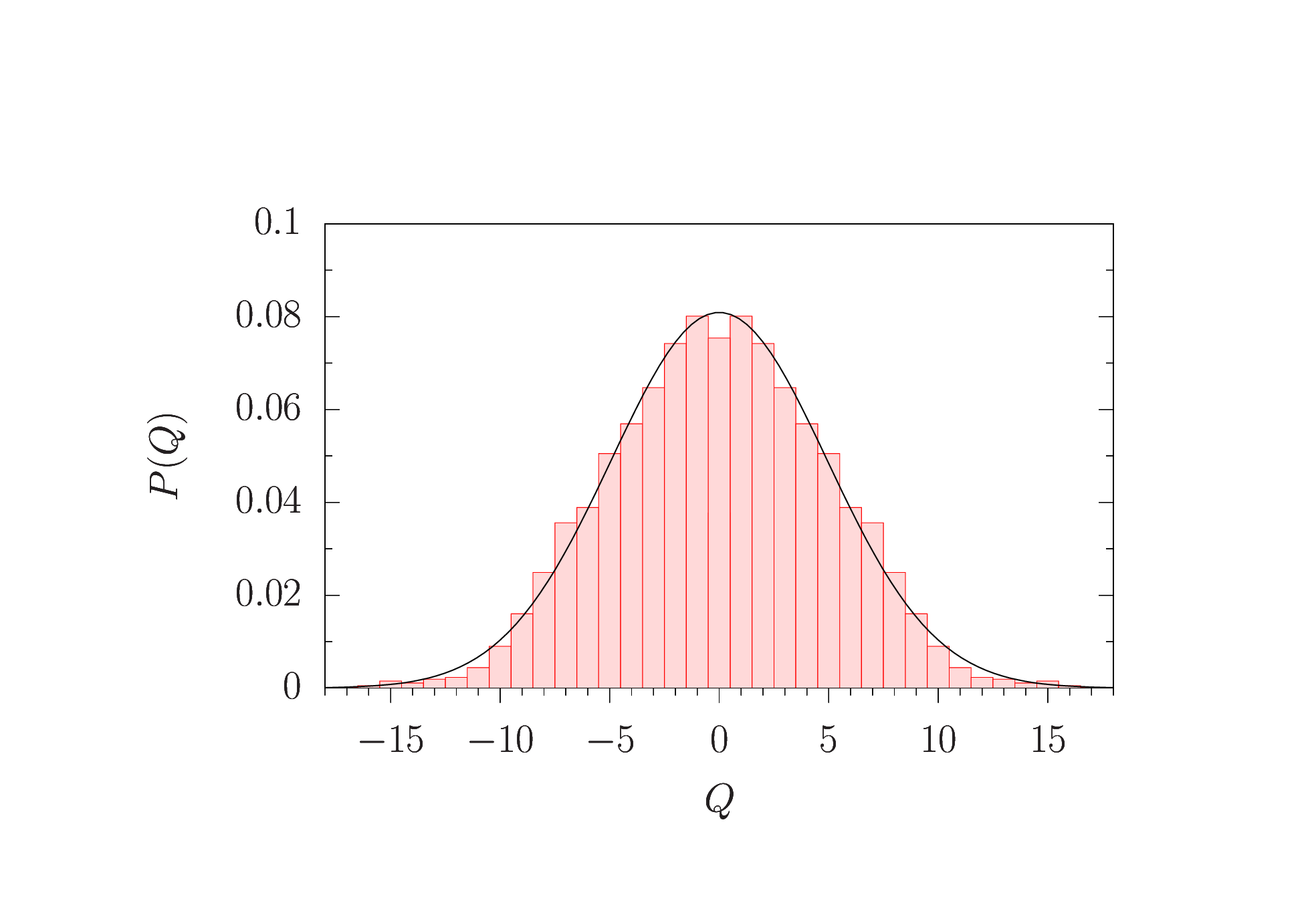,width=7.25cm,clip=} \end{center} \vspace*{-0.9cm}
  \caption{The distribution of topological charge $P(Q)$, together with a Gaussian fit.} \label{fig4}
\end{wrapfigure}
as a function of topological charge $Q$ and flow time $t$ on the $24^4$ lattice. The charge $Q$ has been computed from the field tensor $G_{\mu\nu}$, using the field theoretic definition of the topological charge stated in (\ref{charge}). At large times the action density approaches $E(Q,t) \propto |Q|$, as expected, while the coupling fans out according to $\alpha_V(Q,t) \propto \big(|Q|-\sqrt{(2/\pi)\,\langle Q^2\rangle}\big)\,t^2$. We have checked that the probability distribution of the topological charge, $P(Q)$, is independent of flow time for
$t/a^2 \gtrsim 1$. In Fig.~\ref{fig4} we show $P(Q)$ on the $24^4$ lattice. It turns out that $P(Q)$ is very well described by a Gaussian distribution. The topological susceptibility was found to be $\chi_t=(222(12)\,\mbox{MeV})^4$, which is in the right ballpark.

%
   
The effective coupling constant that describes the interaction of quarks and gluons in the $\theta$ vacuum at the scale $\mu=1/\sqrt{8\,t}$ is given by 
\begin{equation}
  \alpha_V(\theta,t)=\frac{1}{Z_\theta}\int dQ\,e^{i\theta Q} P(Q)\, \alpha_V(Q,t)\,. 
  \label{ft}
\end{equation}
At $\theta=0$ it reduces to $\displaystyle \alpha_V(t)$ as shown in Fig.~\ref{fig2} and by the solid line in Fig.~\ref{fig3} (right panel). We fit $\alpha_V(Q,t)$ by a polynomial in $|Q|$ for each value of $t$ separately. The charge distribution $P(Q)$ is approximated by a Gaussian fitted to the data. The result of the Fourier transform (\ref{ft}) is shown in Fig.~\ref{fig4} for both our lattices. It turns out that $\alpha_V(\theta,t)$ is well approximated by $\displaystyle \alpha_V(\theta,t)/\pi = (\alpha_V(t)/\pi) \big[1-(\alpha_V(t)/\pi)\, (D/\lambda)\,\theta^2\big]^\lambda$, with $\lambda \approx 0.9$ on the $16^4$ lattice and $\lambda \approx 0.75$ on the $24^4$ lattice. At small values of $\alpha_V$ and small flow times $t$, the vacuum angle $\theta$ can assume any value $-\pi \leq \theta \leq +\pi$. Large values of $\alpha_V$, prerequisite for linear confinement, are however only accessible for decreasingly small values of $\theta$. The final result is $\theta=0$ at $\alpha_V(\theta,t)=\infty$. The difference between the curves on the $16^4$ (left panel) and $24^4$ (right panel) lattices, notably for $t/a^2 \gtrsim 40$, can be attributed to finite size effects. Presently it cannot be excluded that at larger $t/a^2$ the parabolas will continue to shrink on larger volumes. 

\begin{figure}[t]
  \begin{center}
    \epsfig{file=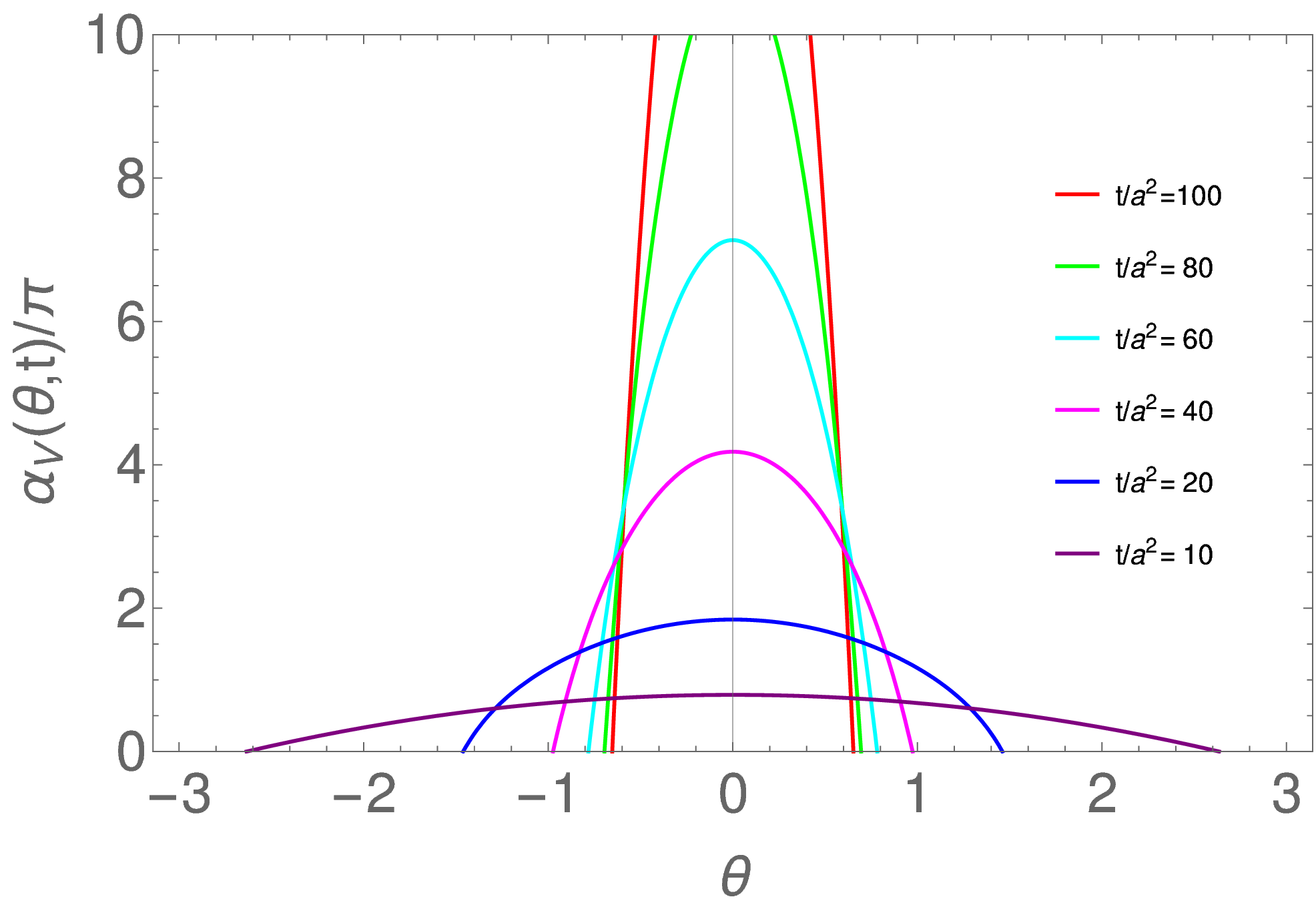,width=7.0cm,clip=}\hspace*{0.25cm}
    \epsfig{file=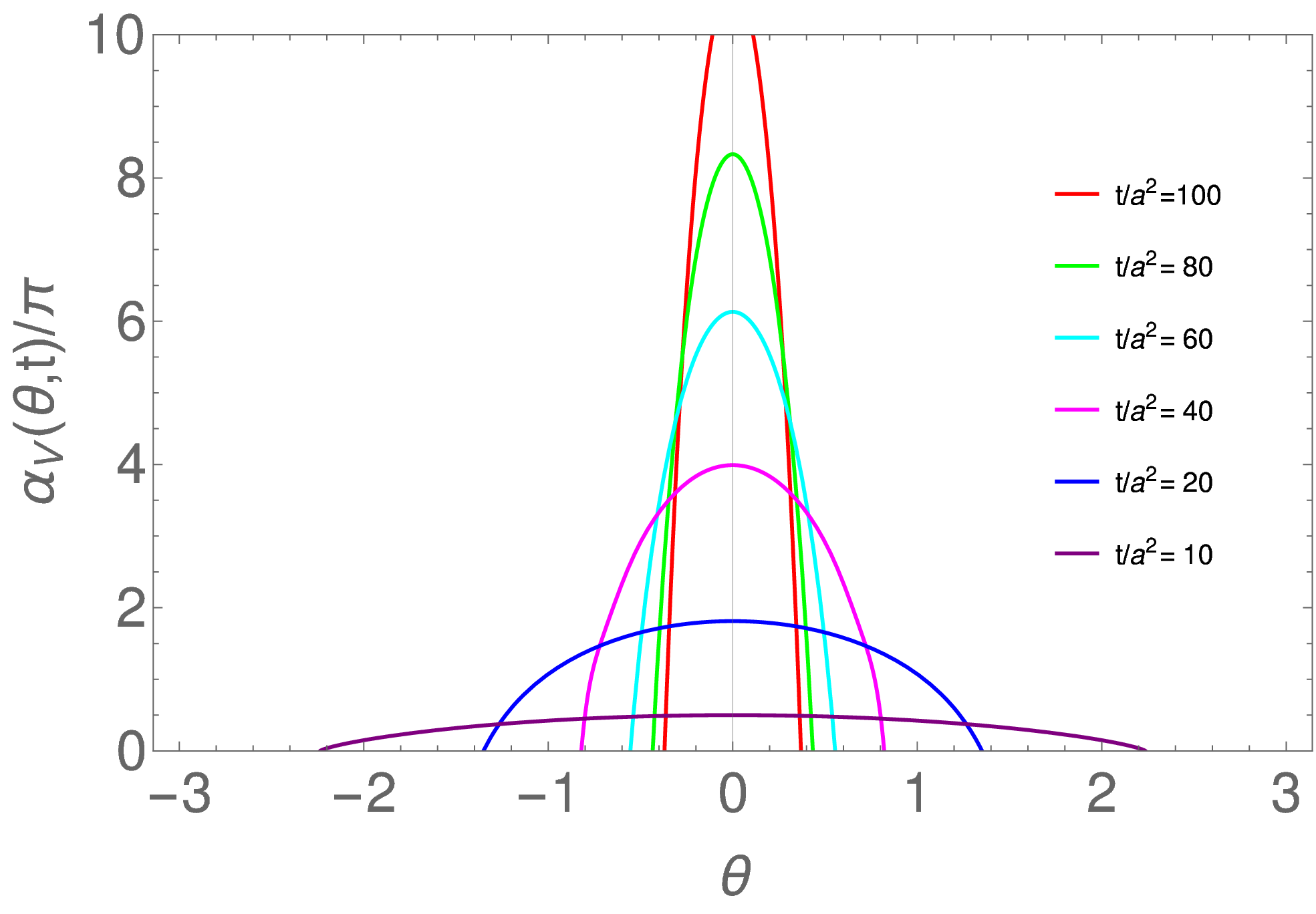,width=7.0cm,clip=}
  \end{center}
  \vspace*{-0.5cm}
  \caption{The running coupling $\alpha_V/\pi$ as a function of $\theta$ for discrete values of flow time $t/a^2$ on the $16^4$ lattice (left panel) and $24^4$ lattice (right panel).}
  \label{fig5}
\end{figure}

From the fitted curves in Fig.~\ref{fig5} we can read off the renormalization group equation for $\alpha_V(\theta,t)$, and the concomitant equation for $\theta(t)$. The appropriate coupling in the infrared regime is $\pi/\alpha_V$. For small values of $\theta$ and $\pi/\alpha_V$ we obtain  
\begin{equation}
  \frac{\partial\, (\pi/\alpha_V)}{\partial \ln\, t} \simeq - \frac{\pi}{\alpha_V} +  D\, \theta^2\,, \quad \frac{\partial\, \theta}{\partial \ln\, t} \simeq - \, \frac{1}{2} \,\theta \,.
  \label{rg}
\end{equation}
Outside this region the equations become increasingly complex. The parameter $D$ turns out to be in good agreement with the expected result, $D = \pi^3 \chi_t/\Lambda_V^4 \approx 1.4$, which follows from approximating the action density $E(t)$ at large $t$ by $|Q|$ times the classical instanton action. The
\begin{wrapfigure}[16]{r}{0.6\linewidth} \begin{center}
  \vspace*{-0.25cm}
    \epsfig{file=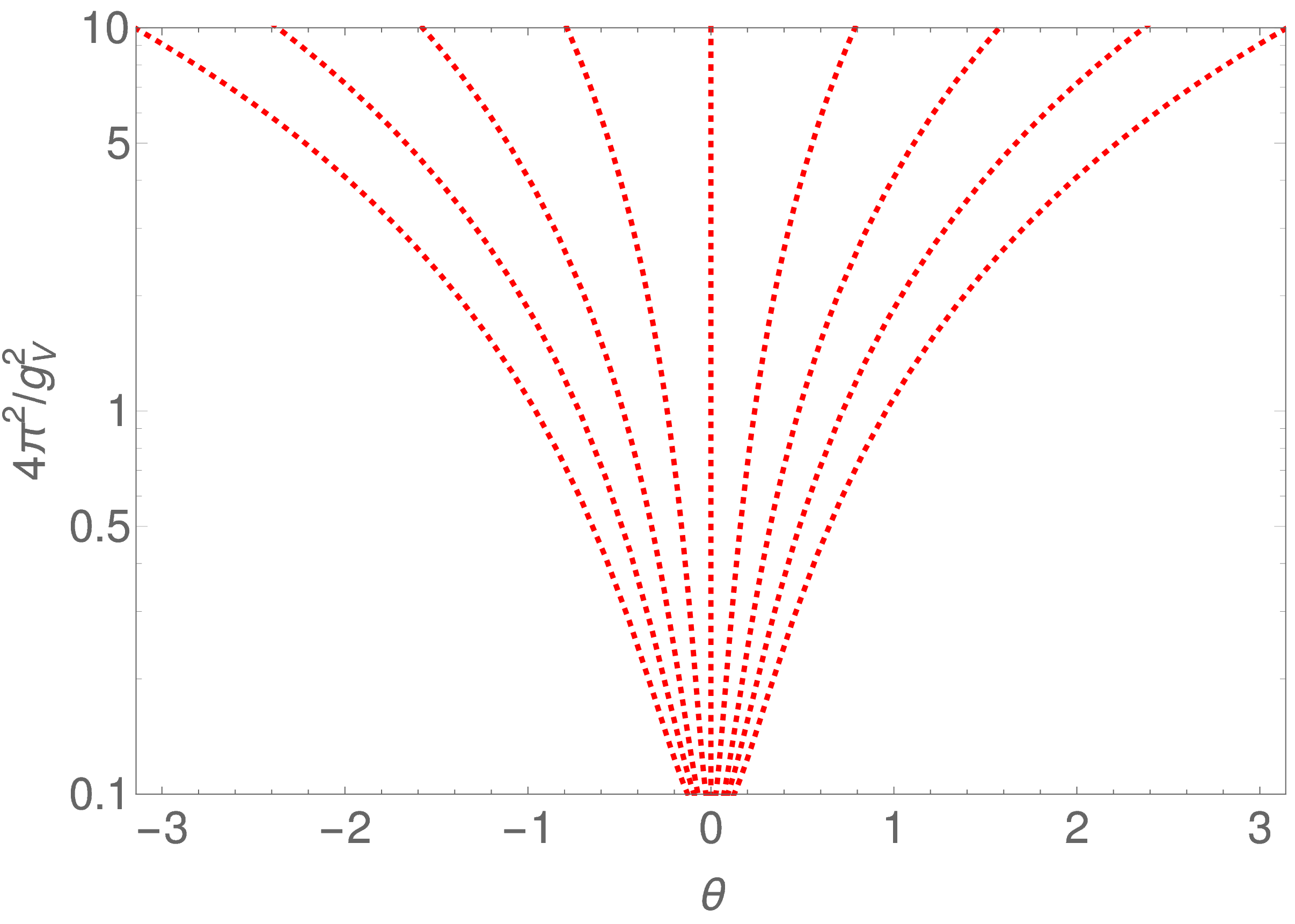,width=8.25cm,clip=}
  \end{center}
  \vspace*{-0.5cm}
  \caption{A logarithmic plot showing the flow of $\theta$ against $4\pi^2/g_V^2$ for different initial values of $\theta$. The result is from the $24^4$ lattice.}
  \label{fig6}
\end{wrapfigure}
renormalization group equations (\ref{rg}) have an infrared attractive, confining fixed point
\begin{equation}
  \frac{\pi}{\alpha_V} = 0\,,\quad \theta = 0 \\[-0.35em]
\end{equation}
at $\mu=1/\sqrt{8\,t}=0$. By numerical integration we obtain the two-parameter renormalization group flow of $\theta$ against $4\pi^2/g_V^2$ sketched in Fig.~\ref{fig6}. The trajectories fill exactly the inner area of the parabolas in Fig.~\ref{fig5}. Fig.~\ref{fig6} shows that any initial value of $\theta$ eventually scales to zero at macroscopic distances, thus providing a mechanism for strong CP conservation. 

\section{Conclusions}

The gradient flow is a powerful tool for studying the long-distance properties of nonabelian gauge theories. In particular, it allows to study the behavior of the theory and its couplings under scale transformations. Motivated by the prospect~\cite{Knizhnik:1984kn} of a dynamical solution of the strong CP problem entirely from QCD, we have studied the renormalization group flow of the running coupling and the vacuum angle $\theta$ in the SU(3) Yang-Mills theory. The main result is that $\theta$ scales to zero in the infrared limit, driven by the confining force and the infrared fixed point at $\pi/\alpha_V=0$. As a result, the $\theta$ term does not lead to any observable effect in the hadronic world. Our calculations indicate that gluonic excitations are largely responsible for this behavior. Quarks, on the other hand, are expected to decouple in the infrared limit and only renormalize $\theta$ at ultraviolet scales.

Clearly, the calculations have to be repeated on larger volumes. We believe that our results are universal, i.e.\ $\displaystyle \alpha_S \stackrel{t\rightarrow\, \infty}{\propto} t$ in any scheme $S$, which needs to be checked in detail as well. Nonetheless, we dare say that there is no theoretical foundation for the Peccei-Quinn model~\cite{Peccei:1977hh} and axions.

\vspace*{-0.15cm}

\section*{Acknowledgement}

GS has been supported in part by DFG grant SCHI 179/8-1. The numerical simulations have been performed at RIKEN Center for Computational Science (Kobe).

\end{document}